\documentclass[lettersize,journal]{IEEEtran}
\usepackage{amsmath,amsfonts}
\usepackage{algorithmic}
\usepackage{algorithm}
\usepackage{array}
\usepackage[caption=false,font=normalsize,labelfont=sf,textfont=sf]{subfig}
\usepackage{textcomp}
\usepackage{stfloats}
\usepackage{url}
\usepackage{verbatim}
\usepackage{graphicx}
\usepackage{cite}
\usepackage{float} 

\usepackage[switch]{lineno}

\begin{document}

\title{HDN: Hybrid Deep-learning and Non-line-of-sight Reconstruction Framework for Transcranial Photoacoustic Imaging of Human Brain}

\author{Pengcheng Wan, Fan Zhang, Yuting Shen, Hulin Zhao, Xiran Cai, \IEEEmembership{Member, IEEE}, Xiaohua Feng, and Fei Gao, \IEEEmembership{Member, IEEE}
\thanks{This work was supported by the National Natural Science Foundation of China under Grant 61805139. (Corresponding authors: Hulin Zhao, Xiran Cai, Xiaohua Feng, and Fei Gao.)}

\thanks{Pengcheng Wan, Fan Zhang, Yuting Shen, and Xiran Cai are with the School of Information Science and Technology, ShanghaiTech University, Shanghai 201210, China (e-mail: wanpc@shanghaitech.edu.cn; zhangfan2022@shanghaitech.edu.cn; 489399125@qq.com; caixr@shanghaitech.edu.cn).}
\thanks{Hulin Zhao is with the Department of Neurosurgery, Chinese PLA General Hospital, Beijing 100853, China.}
\thanks{Xiaohua Feng is with College of Optical Science and Engineering, Zhejiang University, Hangzhou, Zhejiang, 311131, China (e-mail: xiaohuafeng@zju.edu.cn).}
\thanks{Fei Gao is with the Hybrid Imaging System Laboratory, Suzhou Institute for Advanced Research, University of Science and Technology of China, Suzhou, Jiangsu, 215123, China. He is also with the School of Biomedical Engineering, Division of Life Sciences and Medicine, University of Science and Technology of China, Hefei, Anhui, 230026, China, and also with the School of Engineering Science, University of Science and Technology of China, Hefei, Anhui, 230026, China. (e-mail: fgao@ustc.edu.cn).}
}

\markboth{Journal of \LaTeX\ Class Files,~Vol.~14, No.~8, August~2021}%
{Shell \MakeLowercase{\textit{et al.}}: A Sample Article Using IEEEtran.cls for IEEE Journals}

\maketitle

\begin{abstract}
Photoacoustic imaging combines the high contrast of optical imaging with the deep penetration depth of ultrasonic imaging, showing great potential in cerebrovascular disease detection. However, the ultrasonic wave suffers strong attenuation and multi-scattering when it passes through the skull tissue, resulting in the distortion of the collected photoacoustic signal. In this paper, inspired by the principles of deep learning and non-line-of-sight imaging, we propose an image reconstruction framework named HDN (Hybrid Deep-learning and Non-line-of-sight), which consists of the signal extraction part and difference utilization part. The signal extraction part is used to correct the distorted signal and reconstruct an initial image. The difference utilization part is used to make further use of the signal difference between the distorted signal and corrected signal, reconstructing the residual image between the initial image and the target image. The test results on a photoacoustic digital brain simulation dataset show that compared with the traditional method (delay-and-sum) and deep-learning-based method (UNet), the HDN achieved superior performance in both signal correction and image reconstruction. Specifically for the structural similarity index, the HDN reached 0.661 in imaging results, compared to 0.157 for the delay-and-sum method and 0.305 for the deep-learning-based method.
\end{abstract}


\section{Introduction}
Cerebrovascular diseases have become one of the most dangerous fatal diseases in the world, and stroke has been listed as one of the top ten causes of death\cite{donkor2018}. Timely and effective detection of cerebrovascular stenosis or broken can significantly reduce the mortality of patients.

Photoacoustic imaging (PAI) has emerged as a hybrid imaging modality synergistically combining rich optical contrast with high ultrasonic resolution to provide functional and molecular information from deep tissues. In recent years, PAI has been found to have great potential in human brain imaging\cite{na2021}. However, the attenuation and scattering by the skull tissue will cause aberration of the photoacoustic (PA) signal which is a big challenge for brain imaging.

\begin{figure}[ht]
\centering
\includegraphics[width=0.4\textwidth]{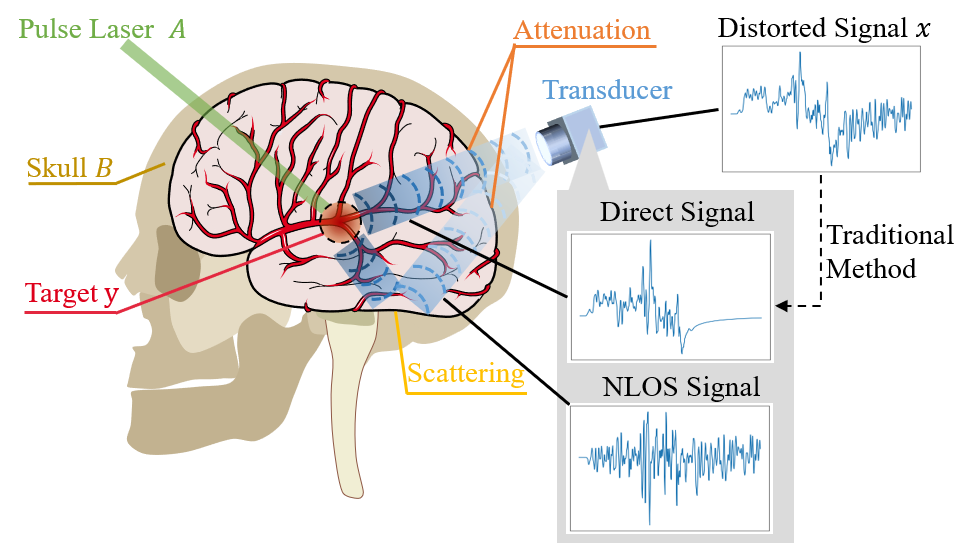}
\caption{Illustration of NLOS PA imaging of vessels in the human brain. Firstly, the pulsed laser hit the blood vessels through the skull. Then the PA signals generated by blood vessels are attenuated and scattered when they encounter the skull. Finally, the transducer receives the distorted signal superimposed by the direct signal (the correction target) and the NLOS signal (discarded as the noise).}\label{nlos}
\end{figure}

\begin{figure*}[h]
\centering
\includegraphics[width=0.9\textwidth]{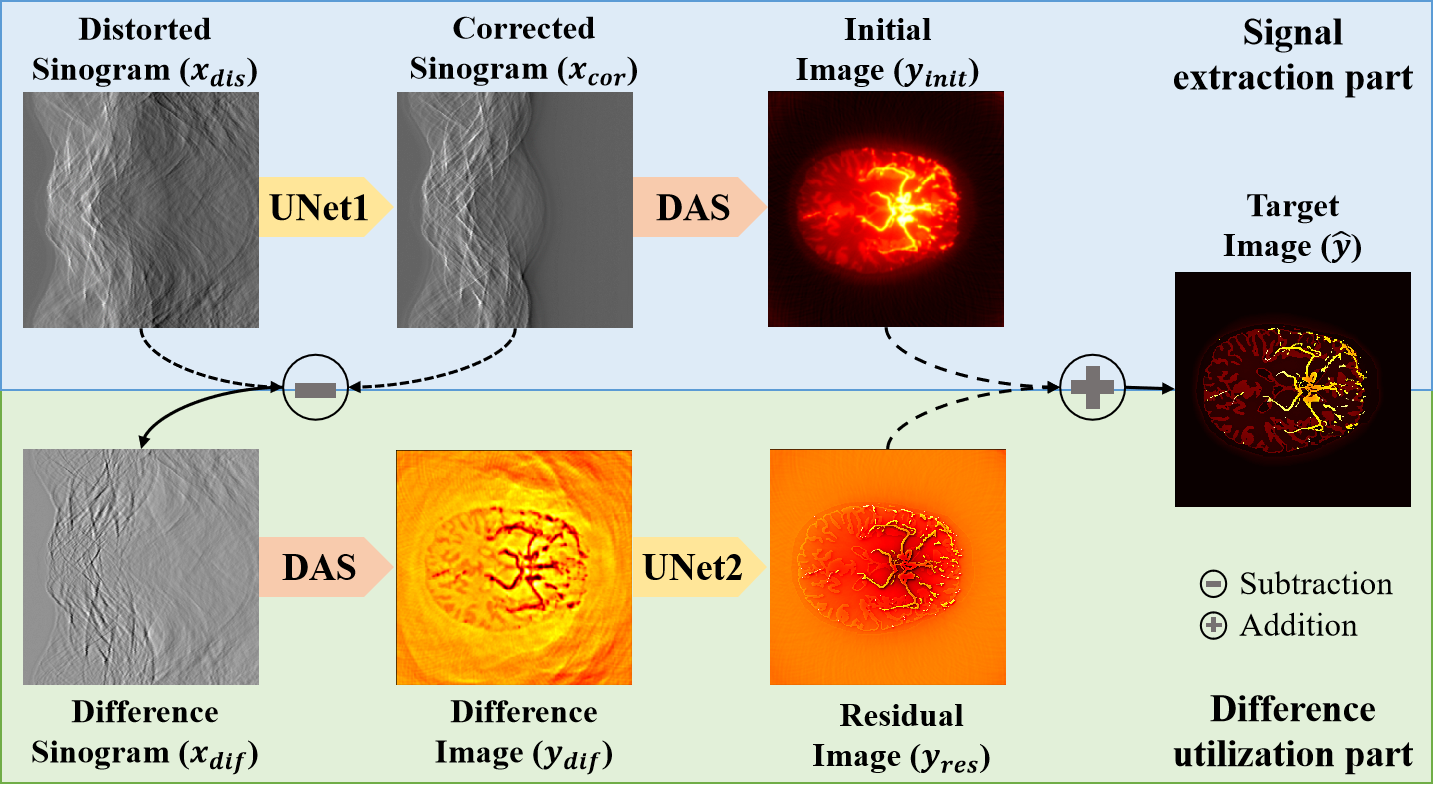}
\caption{The workflow of HDN image reconstruction framework. The arrows represent the method used, and the symbols in the circle represent subtraction or addition operation. All figures have undergone normalization processing, among which the colormap of sinograms and the reconstruction images are respectively the same.}\label{workflow}
\end{figure*}

Chao Huang et al. employ detailed subject-specific descriptions of the acoustic properties of the skull to mitigate skull-induced distortions in the reconstructed image\cite{huang2012}. Hsuan-Kai Huang et al. investigated a learning-based image reconstruction method for three-dimensional (3D) transcranial PACT. The method was systematically assessed in virtual imaging studies that involved stochastic 3D numerical head phantoms and applied to experimental data acquired by use of a physical head phantom that involved a human skull\cite{huang2025}. However, both of this method needs the prior information of CT image data, which are not always available in practice. Liang et al. pointed out that the direct transmission of the shear-wave-converted longitudinal wave is less affected by the skull’s distortion \cite{liang2019,liang2021}. However, in practice, it is challenging to single out the shear wave signal for image reconstruction. Na et al. proposed layered UBP (L-UBP) method, which can de-aberrate the transcranial PA signal by accommodating the skull heterogeneity into conventional UBP. However, the performance in complex craniocerebral mediators remains to be further explored\cite{na2020}. 

In addition, deep-learning-based methods are also popular. Allman et al. proposed the use of convolutional neural networks to identify and remove noise artifacts in photoacoustic signals, which achieved a 100\% success rate in classifying both sources and artifacts\cite{allman2018}. Awasthi et al. use deep neural network for super-resolution, denoising as well as bandwidth enhancement of the PA signals collected at the boundary of the domain\cite{awasthi2020}. Zhang et al. rigorously prepared a photoacoustic digital brain simulation dataset and proposed the use of UNet to correct the PA signal's distortion. The experimental result shows that the skull’s acoustic aberration can be effectively alleviated after UNet correction, achieving conspicuous improvement in PAT human brain images reconstructed from the corrected PA signal, which can clearly show the cerebral artery distribution inside the human skull\cite{zhang2023}. Li et al. propose a polarized self-attention dense U-Net, termed PSAD-UNet, to correct the distortion and accurately recover imaged objects beneath bone plates\cite{li2024}.

However, most of the above-mentioned methods focus on how to identify and remove noise or artifacts, few works noticed that within the so-called "noise" or "artifacts", there may be some potentially useful information. Recently, Shen et al. introduced the non-line-of-sight (NLOS) imaging into PAI, which treated the skull as an intermediate surface and selected the temporal bone as the imaging window. This method uses reflected PA signal by the inner surface of the skull to expand the imaging field of view and further improve the imaging quality\cite{shen2023}.

In this paper, we propose a novel transcranial PAI image reconstruction framework, hybridizing deep-learning with NLOS imaging. The overview of this paper is arranged as follows. Firstly, we briefly introduced the background in section \ref{Background}. Then, in section \ref{Methods}, we describe the proposed method in detail. In section \ref{Simulation} and \ref{Results}, we introduce the simulation detail and present the results of the model. Finally, we draw discussion and conclusion in section \ref{Conclusion}.

\section{Background}
\label{Background}
\subsection{Photoacoustic Imaging}
PAI is based on the PA effect, in which pulsed laser energy is converted into acoustic energy by light absorption. When the laser pulse width meets the thermal confinement and stress confinement, for an ideal point transducer placed at $\vec{r}_d$, the detected PA signal can be written as\cite{xia2014}:
    \begin{equation}
    \label{eq1}
    p_d(\vec{r}_d,t)=\frac{\partial}{\partial t}\left[\frac{t}{4\pi}\iint_{|\vec{r}_d-\vec{r}|=v_st}p_0(\vec{r})d\Omega\right]
    \end{equation}
where, $d\Omega$ is the solid-angle element of $\vec{r}$ with respect to the point at $\vec{r}_d$, $v_s$ is the speed of sound, and $p_0(\vec{r})$ is the initial pressure distribution. Then use $x$ and $y$ to represent the PA signal received by the transducers and the initial pressure distribution, respectively. The forward process of PAI in Eq.\ref{eq1} can be  simplified as:
    \begin{equation}
    \label{eq2}
    x = Ay
    \end{equation}
where $A$ is a linear operator. The goal of PAI is to solve the above inverse problem by recovering $y$ from the known $x$.

Delay-and-sum (DAS) is a common and simple analytic reconstruction method. Its basic idea is to back project the PA signal at different positions in the same phase\cite{ma2020}:
\begin{equation}
    y(i,j) = \sum_{n=1}^N x(n,t(i,j,n))
\end{equation}
where $i\in[1,I],j\in[1,J]$ is the size of imaging area, $N$ is the number of the transducers. $t\in[1,T]$ is the delay time of the propagation time of acoustic wave generated at $(i, j)$ to the $n$th transducer element.

\begin{figure}[h]
\centering
\includegraphics[width=0.5\textwidth]{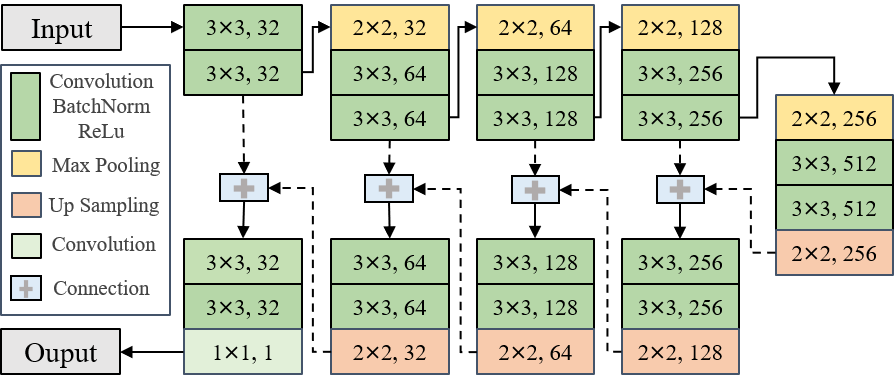}
\caption{The structure of UNet used in this paper. The numbers in the box represent the kernel size and the number of output channels of the this layer, respectively.}\label{unet}
\end{figure}

\subsection{Cerebrovascular Disease Detection}
The clinically available techniques for cerebrovascular diseases include magnetic resonance angiography (MRA), computed tomography angiography (CTA), and digital subtraction angiography (DSA)\cite{liu2024}. Though the above methods can reveal clear vascular structure, they also have some drawbacks and limitations. Among them, MRA has problems such as in-plane saturation, intravoxel dephasing, and long acquisition times, which is not advisable for acute stroke patients\cite{ozsarlak2004}. Although CTA requires less time, it has problems such as radiation exposure and the risk of contrast nephropathy\cite{menon2011}. DSA is the gold standard for diagnostic cerebrovascular assessment, which is routinely used in the context of neuro-interventional and vascular neurology. Still, it also has some disadvantages, such as higher training demands, complicated pre-procedural preparation, and prohibition to poor renal function patients\cite{shaban2022}.

Considering that PAI plays an outstanding role in revealing vasculature, it seems that PAI can be competent in the cerebral vascular diagnosis task\cite{attia2019}. In transcranial PAI, the process of receiving signals under ideal conditions is shown in Eq.\ref{eq2}. However, due to the existence of the skull, the received signal is multi-scattered and superimposed many times, which will lead to the distortion of the signal:
\begin{equation}
\label{eq4}
    x_{dis}(t) = Ay + \sum_{i=0}^{t-1}B_ix(i)
\end{equation}
where $B_i$ represents the reflection or scattering process of the signal $x_i$ caused by the skull. It is important to note that $B_i$ is related to the contact position and the current time $t$ so does not have time invariance, so it is difficult to decompose $x(t)$ into the target $Ay$ and remainder term by the conventional method.

\subsection{Non-line-of-sight Imaging}
NLOS imaging is a kind of computational imaging technology developed in recent years. It collects the light or sound signal $x$ reflected by the intermediate highly-scattering surface through the camera or transducer, to generate the image of the target $y$ beyond the line of sight. Currently, NLOS has been widely used in security monitoring, rescue missions, medical imaging, and other fields\cite{maeda2019b}.

The imaging process of NLOS is similar to the Eq.\ref{eq4}, however, due to the obstruction, the signal $Ay$ transmitted by the target cannot be directly received. We receive the signal by multiple reflection, scattering from the intermediate plane instead:
\begin{equation}
\label{eq5}
x_{dif}(t) = B_tAy + \sum_{i=0}^{t-1}B_ix(i)
\end{equation}
this is similar to the second item in the Eq.\ref{eq4}. As shown in Fig.\ref{nlos}, if we treat the skull as an intermediate surface and treat the received signal (in Eq.\ref{eq4}) as the sum of the target signal (in Eq.\ref{eq2}) and the difference signal (in Eq.\ref{eq5}), we hypothesize that a hybrid utilization of both direct signal and difference signal will help to improve the quality of the reconstructed image.

\begin{figure}[ht]
\centering
\includegraphics[width=0.5\textwidth]{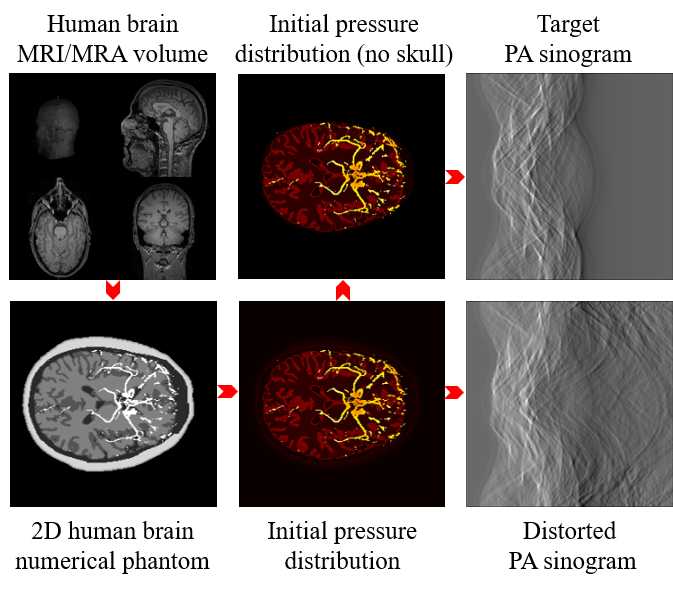}
\caption{The preparation process of PA digital brain dataset\cite{zhang2023}.}\label{dataflow}
\end{figure}

\section{Methods}
\label{Methods}
As shown in Fig.\ref{workflow}, we propose an image reconstruction framework named HDN inspired by NLOS technology. It consists of the signal extraction part and the difference utilization part. A detailed description of the proposed method will be introduced in the following sections.

\begin{figure*}[h]
\centering
\includegraphics[width=0.8\textwidth]{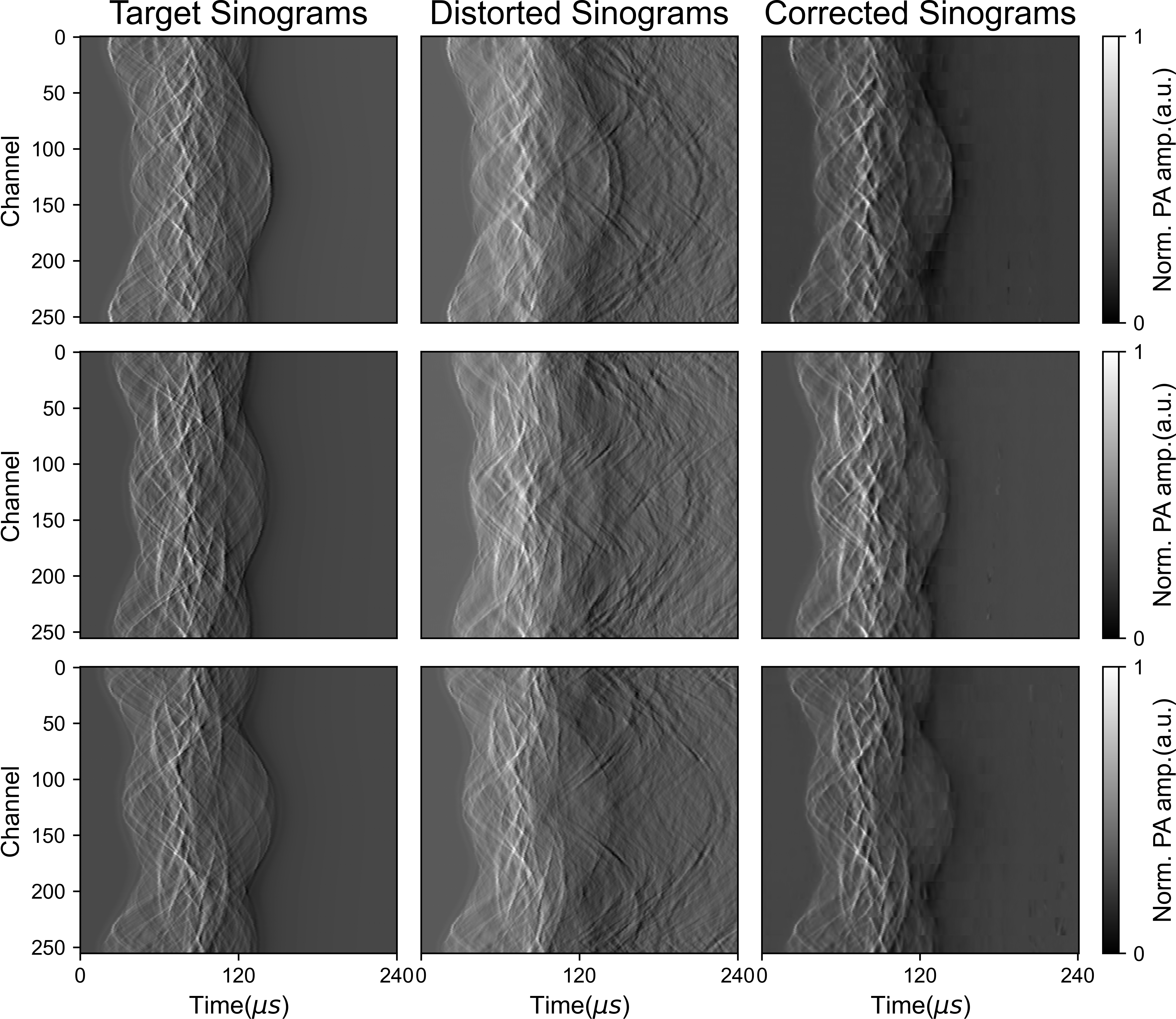}
\caption{Three samples of normalized PA sinograms. The first column shows the target PA sinograms. The second column is the distorted sinograms used for the inputs of HDN. The UNet1 corrected PA sinograms are in the last column.}\label{signal}
\end{figure*}

\subsection{Signal Extraction Part}
In transcranial PA imaging, our primary targets are cerebral vessels. However, the acoustic wavelength of the PA signal generated by microvessels is comparable to the size of skull pores, resulting in Mie scattering rather than Rayleigh scattering. This scattering mode will generate a side-lobe signal, and the superimposition of the main lobe and side lobes complicates the PA image reconstruction process\cite{gao2022}.

To alleviate PA signal distortion due to the skull tissue, in the signal extraction part, we propose to use the deep-learning based method to extract a clean signal from the distorted signal and reconstruct the initial image. More specifically, we notice that the collected distorted signal ($x_{dis}$) consists of the target signal ($x_{tar}$) and the difference signal ($x_{dif}$):
\begin{equation}
    x_{dis} = x_{tar} + x_{dif}
\end{equation}
Therefore, we first use the neural network model to extract the corrected signal from the distorted signal:
\begin{align}
    x_{cor} &= \mathrm{U}_1(\theta_1^*, x_{dis})\\
    \theta_1^* &= \mathop{\arg\min}\limits_{\theta_1} \left \| x_{tar} - \mathrm{U}_1(\theta_1, x_{dis}) \right \|_{2}^{2}
\end{align}
where $\mathrm{U}(\theta,\cdot)$ denote the neural network map with parameter $\theta$. Then we use the DAS algorithm to reconstruct the initial image. So in the signal extraction part, our goal is to get the initial image from the distortion signal:
\begin{equation}
    y_{init} = \mathrm{DAS}(\mathrm{U}_1(\theta_1^*,x_{dis}))
\end{equation}

\begin{figure*}[h]
\centering
\includegraphics[width=0.8\textwidth]{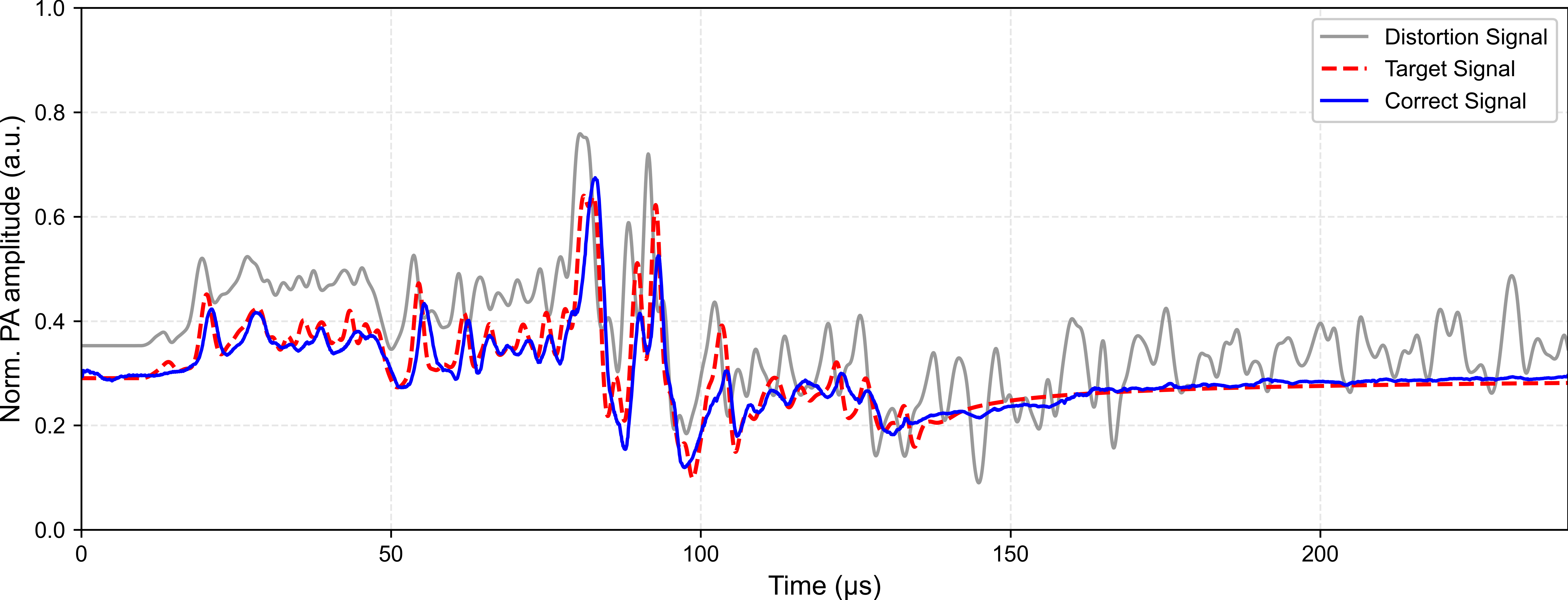}
\caption{Different result of normalized first channel of PA signal from the last smaple of Fig.\ref{signal}. }\label{channel_signal}
\end{figure*}

\subsection{Difference Utilization Part}
Previous work only focuses on extracting the corrected signal from the distorted signal. However, it deserves noting that these removed distortions still contain some useful physical information. Inspired by NLOS technology, we further add the difference utilization part to utilize the removed difference signal. Specifically, we treat the skull as the reflecting surface, and named the difference between the distorted signal and the corrected signal as the difference signal to further use it. 

It is worth noting that the difference signal experienced multiple reflections and scattering, which makes it difficult to extract features from the signal domain. Therefore, different from the signal extraction part, we first use the DAS algorithm to reconstruct the difference image:
\begin{equation}
    y_{dif} = \mathrm{DAS}(x_{dis}-x_{cor})
\end{equation}
After transferring the features from the signal domain to the image domain, then still use neural network $U_2$ to learn the residual between the initial image and the target image:
\begin{align}
    y_{res} &= \mathrm{U}_2(\theta_2^*, y_{dif})\\
    \theta_2^* &= \mathop{\arg\min}\limits_{\theta_2} \left\| (y-y_{init}) - \mathrm{U}_2(\theta_2, y_{dif}) \right\|_{2}^{2}
\end{align}
By adding the residual image and the initial image, the final imaging result can be obtained:
\begin{equation}
\begin{aligned}
    \hat{y} &= y_{init} + y_{res}\\
            &=\mathrm{DAS}(\mathrm{U}_1(\theta_1^*, x_{dis}))+\mathrm{U}_2(\theta_2^*, \mathrm{DAS}(x_{dis}-\mathrm{U}_1(\theta_1^*, x_{dis}))) 
\end{aligned}
\end{equation}

\subsection{Convolutional neural network}
In both the signal extraction part and the difference utilization part, we need to use neural networks to complete the feature extraction work within the same domain. Here, we choose to use the classic convolutional neural network UNet to complete this task, the UNet\cite{ronneberger2015} structure used in this work is shown in Fig.\ref{unet}. More precisely, we use two UNet with exactly the same structure but different parameters, namely $U_1$ and $U_2$. Among them, $U_1$ is used to extract the correction signal from the distorted signal, and $U_2$ is used to extract the complementary information of the initial image from the difference image, that is, the residual image.

We choose UNet for the following reasons. Firstly, due to the distorted signal and the corrected signal, as well as the difference image and the residual image, are all relatively similar, thus the UNet is highly suitable for the task of this paper, as a typical convolutional neural network with residual structure. Furthermore, the residual structure can fully extract the effective information while ensuring the main features remain unchanged. In addition, its special encoding and decoding structure can fully extract the potential information in the data, which is very suitable for the situation of small data set.

\section{Simulation Detail}
\label{Simulation}
\subsection{Photoacoustic Digital Brain Dataset}
The deep-learning-based approach is a data-driven approach that requires large amounts of data to be trained to get the desired results. However, PAI as a newly developed imaging technology, has not been able to obtain a large amount of clinical data to train neural networks. Therefore, we trained and tested the HDN using the PA digital brain dataset presented in \cite{zhang2023}.

This dataset has a rigorous preparation process. Firstly, T1-weighted 3D MRA images from the IXI dataset from Hammersmith Hospital were used to generate a 2D numerical model of the human brain. By superimposing the six types of tissue of the scalp, skull, vessel, gray matter, white matter, and cerebrospinal fluid, it can generate a pseudo-3D human brain numerical model size of $12\times256\times256$. More specifically, from top to bottom, it consists of two layers of scalp, five layers of scalp and skull mixture layer, four layers of scalp, skull, and cerebrospinal fluid mixture layer, and one layer of all six types of tissue mixture layer mentioned above. Then, the optical simulation of the human brain numerical model was carried out by MCXLAB\cite{fang2009}, obtaining corresponding optical fluence. The initial pressure distribution is obtained by multiplying the optical flux with the corresponding optical absorption coefficient. Finally, the initial pressure distribution and its removal of skull and scalp tissue were respectively imported into the k-Wave toolbox for acoustic simulation\cite{treeby2010}, generating distorted signal and target signal. The whole data preparation process is shown in Fig.\ref{dataflow}.

\begin{figure*}[t]
\centering
\includegraphics[width=0.8\textwidth]{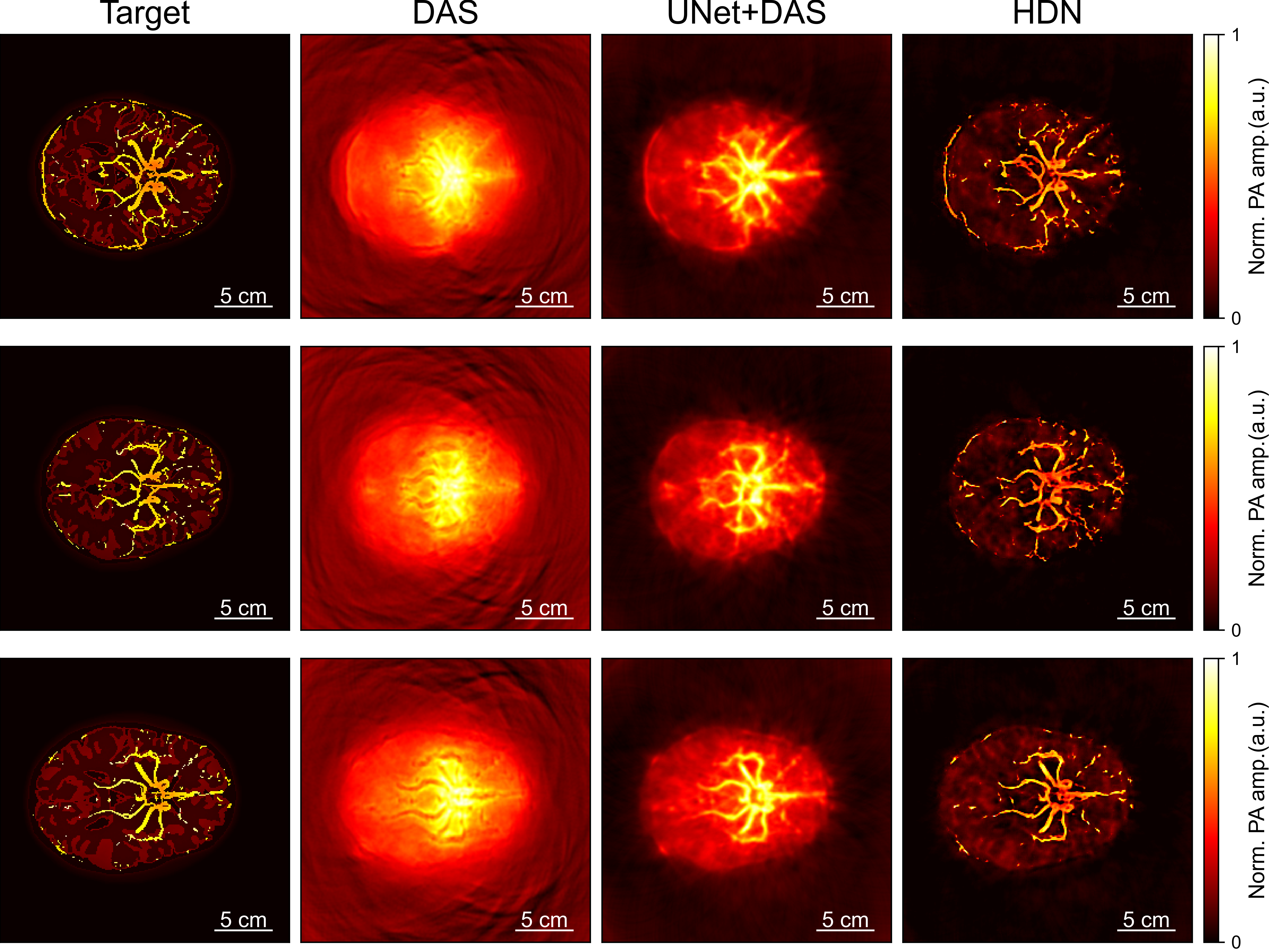}
\caption{Three samples of normalized imaging results. The first column is the initial pressure distribution, which is the imaging target. The second column is the imaging results of the distorted signal directly reconstructed by DAS. The third column is the initial images in the signal extraction part. The last column is the HDN imaging results.}\label{image}
\end{figure*}
    
\subsection{Model Training Details}
We trained the HDN using the dataset presented in \cite{zhang2023}, which contains 180 sets of $256\times3001$ distorted signal, target signal, and the corresponding $256\times256$ initial pressure distribution. To maintain dimensional consistency after each pooling operation in UNet, we extend the signal to $256\times3008$ by adding zeros to the end of each channel. We randomly divided the data into 144 training sets, 18 validation sets, and 18 test sets.

In order to better enable the network to learn various features, we carry out data augmentation processing on the training sets. Since the transducers in the acoustic simulation are set to a circular shape and each channel signal comes from the corresponding transducer, we divide the signal evenly into 8 segments, each containing 32 channels. We then select 16 channels from each segment in turn and connect the selected data to get a $128\times3008$ sampled signal. By overlapping the channels in each segment, we can get 17 different combinations of sampled signals.

We used the Pytorch deep-learning framework to simultaneously train UNet1 and UNet2 on a NVIDIA A40 GPU with 48GB memory. The iteration is set to 300 epochs and batch size was 1 due to the data augmentation. The optimizers of the two neural networks were Adam, and the learning rates were 0.001. The loss function is the mean square error (MSE), which is calculated by the following formula:
\begin{equation}
\label{mse}
\mathrm{MSE}(y, \hat{y}) = \frac{1}{n}\sum_{i=1}^{n}(y_{i}-\hat{y_{i}})^{2}
\end{equation}
where $y$ represents the truth value and $\hat{y}$ represents the output of the neural network. In the UNet1, we calculate the MSE between $x_{tar}$ and $x_{cor}$, and in the UNet2, we calculate the MSE between $(y-y_{init})$ and $y_{res}$.

\begin{table}
    \centering
    \caption{Quantitative evaluation of signal correction results.\label{tab1}}
    \begin{tabular}{clll}
    \hline
    & SNR  & MSE  \\ \hline
    \begin{tabular}[c]{@{}l@{}}Distorted signal\end{tabular} & 3.011 & 0.2320 \\
    \begin{tabular}[c]{@{}l@{}}Corrected signal\end{tabular} & 16.93 & 0.0026\\ 
    \hline
    \end{tabular}
\end{table}

\begin{table}
    \centering
    \caption{Quantitative evaluation of image reconstruction results.\label{tab2}}
    \begin{tabular}{clll}
    \hline
    & SSIM  & PSNR    \\ \hline
    DAS                                      & 0.157 & 11.641\\
    UNet1+DAS                                & 0.305 & 15.567\\
    HDN                                      & 0.661 & 21.342\\ \hline
    \end{tabular}
\end{table}

\subsection{Evaluation Metrics}
To better evaluate the performance of our framework, we adopted different evaluation metrics for signal correction and image reconstruction. Specifically, we employed the Signal-to-Noise Ratio (SNR) and MSE (defined in Eq. \eqref{mse}) to evaluate the performance of signal correction. Where SNR measures the preservation of target signal power relative to residual noise (in dB), with higher values indicating cleaner signal recovery, and MSE quantifies the pixel-wise deviation between the target and corrected signals, where lower values denote improved accuracy. The SNR is defined as:
\begin{equation}
    \mathrm{SNR}(x_{tar},x_{cor}) = 10\cdot \log_{10}\left( \frac{\sum^m(x_{tar}^2)/m}{\mathrm{MSE}(x_{tar},x_{cor})} \right)
\end{equation}
and we employed the Structural Similarity Index (SSIM)\cite{wang2004} and Peak Signal-to-Noise Ratio (PSNR) to evaluate the performance of image reconstruction. Where SSIM quantifies the perceptual similarity between two images by comparing luminance, contrast, and structural patterns, with values closer to 1 indicating higher fidelity, and the PSNR measures reconstruction accuracy by computing the logarithmic ratio of the maximum possible pixel intensity to the MSE between images, where higher values (in dB) denote better quality. The are defined as:
\begin{equation}
    \mathrm{SSIM}(y,\hat{y})=\frac{(2\mu_{y}\mu_{\hat{y}}+C_{1})(2\sigma_{y\hat{y}}+C_{2})}{(\mu_{y}^2+\mu_{\hat{y}}^2+C_{1})(\sigma_{y}^2+\sigma_{\hat{y}}^2+C_{2})}
\end{equation}

\begin{equation}
    \mathrm{PSNR}(y,\hat{y})=10\cdot \log_{10}\left(\frac{I^2}{\mathrm{MSE}(y,\hat{y})} \right)
\end{equation}
where $I$ is the maximum value of image, $C_1=(0.01I)^2,C_2=(0.03I)^2$, $\mu_y$ denote the mean vaule of $y$, and $\sigma_y$ denote the variance of $y$.

\begin{figure*}[t]
\centering
\includegraphics[width=0.8\textwidth]{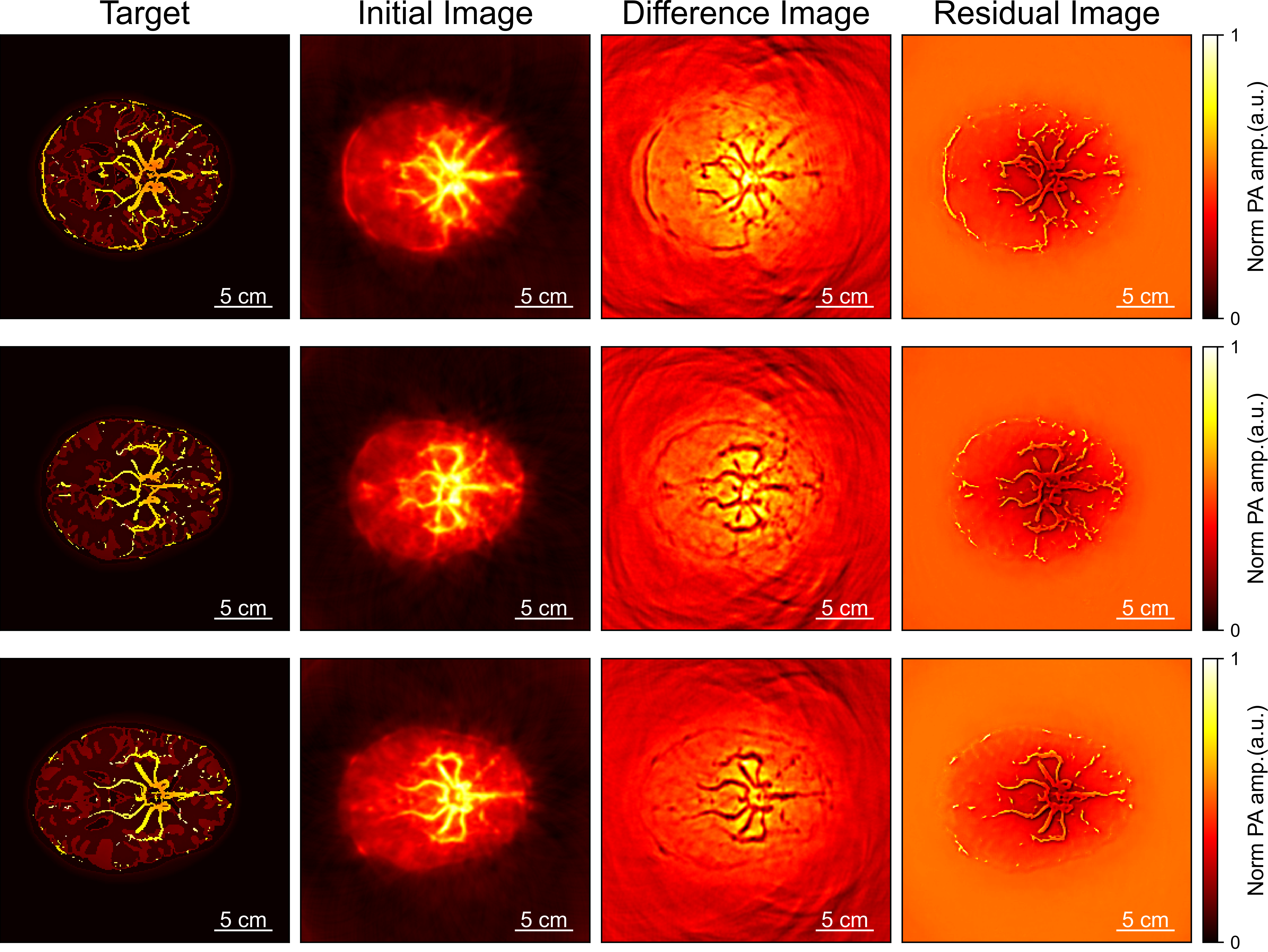}
\caption{Three samples of normalized intermediate image in the HDN framework. The first column is the imaging target. The second column is the initial image in the signal extraction part. The last two columns are the difference image and residual image respectively in the difference utilization part.}\label{ablation}
\end{figure*}

\section{Results}
\label{Results}
We did the same data augmentation for the test sets and tested them on the trained model, getting the output and doing the corresponding inverse transformation.

Three samples of normalized target PA sinograms, distorted PA sinograms, and UNet1 corrected PA sinograms are shown in Fig.\ref{signal}. It shows that the UNet1 effectively separates the clean PA sinograms from the distorted PA sinograms, although there is a certain lack of detail (after $120 \mu s$), this does not greatly affect the imaging quality. The normalized PA signals taken from the first channel of PA sinograms acquired from the third object in Fig.\ref{signal} in different status are shown in Fig.\ref{channel_signal}, in which we can see that the corrected signal is closer to the signal target in both phase and waveform compared to the distorted signal.

The imaging results of the same three samples are shown in Fig.\ref{image}, compared with the traditional method (DAS) and deep-learning-based method\cite{zhang2023} (UNet+DAS, thus the initial image), the HDN has achieved obvious improvement in both image contrast and resolution of cerebrovascular imaging, especially in tiny blood vessels. There is even a slight improvement in the distinction between gray/white matter. Furthermore, the difference between the initial image and the HDN results demonstrated the importance of the difference utilization part, validating the value of multi-reflected PA signals, which are conventionally regarded as "noise" or "artifacts".

To better quantify the results, we calculate the SNR and MSE evaluation indexes between the target signal and the distorted signal before and after signal correction, and calculate the SSIM and PSNR evaluation indexes between the target image and the imaging results by different methods. As shown in Tab.\ref{tab1}, we found that after correction, the signal has been improved in both the evaluation indexes, especially in the SSIM. Similarly, as shown in Tab.\ref{tab2}, in terms of imaging results, compared with the traditional method (DAS) and deep-learning-based method, the HDN greatly improves imaging quality in both evaluation indexes.

\section{Discussion and Conclusion}
\label{Conclusion}
In this paper, we propose a transcranial PA image reconstruction framework named HDN, which consists of the signal extraction part and the difference utilization part. The signal extraction part corrects the distorted signal and reconstructs the initial image. The difference utilization part is used to make further use of the signal difference and reconstruct the residual image between the initial image and the target image.

In signal extraction part, it is a common step to denoise the PA signal and then reconstruct the initial image in most PA imaging tasks. But inspired by NLOS, we thought that there might still be useful information in the discarded signals, and thus further added the part of difference utilization, where we get the difference signal by subtracting the distorted signal and the corrected signal. However, considering that the difference signal contains superposition of multiple reflections, scatterings, and noise, it is necessary to map the difference signal to the image domain for feature extraction. The difference image shows the information existing in the difference signal very well as shown in the third column of Fig.\ref{ablation}.

Considering that DAS algorithm is a kind of linear transformation, we do not directly add the initial image to the difference image. In fact, it is no different from imaging directly using DAS. This will lead to a worse result than the initial image as shown in the second column in Fig.\ref{image}. So to make better use of the information in the difference image, we use UNet2 to learn the residual map between the initial image and the target as a kind of compensation. 

As shown in last column in Fig.\ref{ablation}, since the training of unet2 uses the prior information of the initial image and the target image, and the initial image is directly derived from the result of the signal extraction part instead of reconstructing the true value of the target signal using DAS, the residual image shows a clearer vascular structure compared to the initial image. Therefore, the residual image is not only the complementarity of the initial image, but also the feature extraction of the difference image and the reuse of the initial image.

Finally, the HDN is tested on the PA digital brain simulation data set, and the test results show that compared with the traditional method and deep-learning-based method, the HDN has good performance in both signal correction and image reconstruction. Future work is to further validate the HDN in the clinic data and consider other better alternative algorithms for signal corrects and image reconstruction.

\section{Acknowledgment}
This work received equipment support from the Core Facility Platform of Electronics and
computation support from the HPC Platform of ShanghaiTech University.

\bibliographystyle{IEEEtran}
\bibliography{reference}

\end{document}